\documentclass[doublecol]{epl2}
\usepackage{amsmath}

\newcommand{\dd}{\text{d}}
\newcommand{\ee}{\text{e}}

\newcommand{\alp}{\alpha}
\newcommand{\ta}{\tau_\alp}
\newcommand{\kb}{k_\text{\tiny B}}
\newcommand{\va}{v_\text{\tiny A}}
\newcommand{\Ta}{T_\text{\tiny A}}
\newcommand{\Tf}{T_\text{eff}}

\newcommand{\tc}{\tilde{\chi}}
\newcommand{\tC}{\tilde{C}}

\newcommand{\tz}{\tilde{\zeta}}

\newcommand{\fig}{Fig.}

\providecommand{\avg}[1]{\left \langle #1 \right \rangle}
\providecommand{\pnt}[1]{\left ( #1 \right)}
\providecommand{\brt}[1]{\left [ #1 \right]}
\providecommand{\abs}[1]{\left | #1 \right|}
\providecommand{\f}[2]{\frac{ #1}{#2}}

\title{Nonequilibrium dissipation in living oocytes}

\author{\'E. Fodor\inst{1,*} \and W. W. Ahmed\inst{2,3*} \and M. Almonacid\inst{4,*} \and M. Bussonnier\inst{2} \and N. S. Gov\inst{5} \and M.-H. Verlhac\inst{4} \and T. Betz\inst{2} \and P. Visco \inst{1} \and F. van Wijland\inst{1,6}}
\shortauthor{\'E. Fodor \etal}

\institute{                    
  \inst{1} Laboratoire Mati\`ere et Syst\`emes Complexes, UMR 7057 CNRS/P7, Universit\'e Paris Diderot, 10 rue
  Alice Domon et L\'eonie Duquet, 75205 Paris cedex 13, France
	\\
  \inst{2} Laboratoire Physico-Chimie Curie, Institut Curie, PSL Research University, CNRS UMR168, 75005, Paris, France; Sorbonne Universit\'es, UPMC Univ Paris 06, 75005, Paris, France
	\\
	\inst{3} Department of Physics, California State University, Fullerton, California 92831, USA
	\\
  \inst{4} CIRB, Coll\`ege de France, CNRS-UMR7241, INSERM-U1050, 75231 Paris, Cedex 05, France
	\\
	\inst{5} Department of Chemical Physics, Weizmann Institute of Science, 76100 Rehovot, Israel
	\\
	\inst{6} Department of Chemistry, University of California, Berkeley, CA, 94720, USA
	\\
	\inst{*} These authors contributed equally to this work
}


\abstract{
	Living organisms are inherently out-of-equilibrium systems. We employ recent developments in stochastic energetics and rely on a minimal microscopic model to predict the amount of mechanical energy dissipated by such dynamics. Our model includes complex rheological effects and nonequilibrium stochastic forces. By performing active microrheology and tracking micron-sized vesicles in the cytoplasm of living oocytes, we provide unprecedented measurements of the spectrum of  dissipated energy. We show that our model is fully consistent with  the experimental data, and we use it to offer predictions for the injection and dissipation energy scales involved in active fluctuations.
}

\begin{document}

\maketitle


Perrin's century old picture~\cite{perrin1908origine} where the Brownian motion of a colloid results from the many collisions exerted by the solvent's molecules is a cornerstone of soft-matter physics. Langevin~\cite{langevin_sur_1908} modeled the ensuing energy exchanges between the solvent and the colloidal particle in terms of a dissipation channel and energy injection kicks. The key ingredient in the success of that theory was to completely integrate out the "uninteresting" degrees of freedom of the solvent whose properties are  gathered in a friction constant and a temperature. In this work we take exactly the reverse stance and ask how, by observing the motion of a tracer embedded in a living medium, one can infer the amount of energy exchange and dissipation with the surrounding medium. The main goal is to quantify the energetic properties of the medium, both injection and dissipation-wise.

This is a stimulating question because there are of course striking differences between a living cell and its equilibrium polymer gel counterpart, to which newly developed~\cite{Sekimoto:1339587,0034-4885-75-12-126001} methods of nonequilibrium statistical mechanics apply. Beyond thermal exchanges that fall within the scope of a Langevin approach, ATP consumption fuels molecular motor activity and drives relentless rearrangement of the cytoskeleton. This chemically driven continuous injection and dissipation of energy adds a nonequilibrium channel that eludes straightforward quantitative analysis. In short, a living cell is not only a fertile playground for testing new ideas from nonequilibrium physics, but also one in which these ideas can lead to a quantitative evaluation of an otherwise ill-understood activity which is of intrinsic biophysical interest. Our work addresses both aspects by a combination of active microrheology, tracking experiments, and theoretical modeling.

One experimental way to access nonequilibrium physics in the intracellular medium is to focus on the deviation from thermal equilibrium behavior of the tracer's position statistics: forming the ratio of the response of the tracer's position to an infinitesimal external perturbation to its unperturbed mean-square displacement leads to a quantity that only reduces to the inverse temperature when in equilibrium, by virtue of the fluctuation-dissipation theorem (FDT). Earlier tracking experiments supplemented by microrheology techniques have allowed the departure from equilibrium to be analyzed in terms of this ratio in a variety of contexts~\cite{Mizuno, Martin:01, GalletSM, Nir, Joanny, Betz} ranging from reconstituted actin gels to single cells. However the limitations inherent to this effective temperature are well-known: it bears no universal meaning as it depends on the observable under scrutiny, thus it cannot be equated to a {\it bona fide} temperature, and hence it does not connect to the underlying microscopic dynamics. In spite of these caveats, the effective temperature has been widely measured in nonequilibrium systems since it is perhaps the simplest way to assess deviation from equilibrium.

Here we exploit a body of theoretical methods that have been developed over the last ten years to infer quantitative information about the nonequilibrium processes driving  intracellular dynamics. Within the realm of stochastic thermodynamics~\cite{Sekimoto:1339587,0034-4885-75-12-126001} --as it strives to extend concepts of macroscopic thermodynamics to small and highly fluctuating systems~\cite{Tusch, Muy, Shinkai}, the Harada-Sasa equality stands out as being particularly suited to our goal~\cite{HarSas,harada2006}. Nonequilibrium systems are characterized by the dissipation of energy, which is absorbed by the surrounding thermostat \textit{via} a transfer from the system to the bath. The Harada-Sasa equality connects the rate of dissipated energy to the spatial fluctuations in a nonequilibrium steady-state system. The feasibility of measuring the various ingredients in the Harada-Sasa framework was demonstrated in model systems such as a micron-sized colloidal particle in a viscous fluid~\cite{PhysRevE.75.011122, ExpAll}, and then later generalized to a viscoelastic medium~\cite{PhysRevE.74.026112}. It has also been used to quantify the efficiency of an isolated molecular motor~\cite{PhysRevLett.104.198103}.

The systems to which we apply this equality are micron-sized vesicles that are present in the cytoplasm of mouse unfertilized eggs, known as oocytes. Their motion in the cell is mainly regulated by myosin-V motors on the actin network~\cite{Howard, Schuh, Almonacid}. The use of such vesicles allows us to capture the intrinsic intracellular dynamics without using artificial external particles that may alter the environment. From a physics perspective, oocytes are also major assets since they constitute a rare example of a living cell that remains steady on the timescales of hours. They are spherical in shape, with typical radius of about $40$~$\mu$m, and their nucleus is centrally located at the end of Prophase I~\cite{Verlhac:13}.

In this paper, we directly access nonequilibrium dissipation within the cell. We first characterize the intrinsic rheology of the medium experienced by the vesicles. Then, we present a minimal microscopic model for the dynamics of the vesicles which is driven by the nonequilibrium reorganization of the cytoskeleton by molecular motor generated forces. Our first main result is to demonstrate that this model is sufficient to capture the available experimental data. As a second step, we use the Harada-Sasa equality to predict and quantify the rate of nonequilibrium dissipated energy in our experimental system. Finally, we employ this new prediction to evaluate how nonequilibrium activity varies across the cell, and we offer interpretations about the role of molecular motors in vesicle motion.

\begin{figure}
\includegraphics[width=\columnwidth]{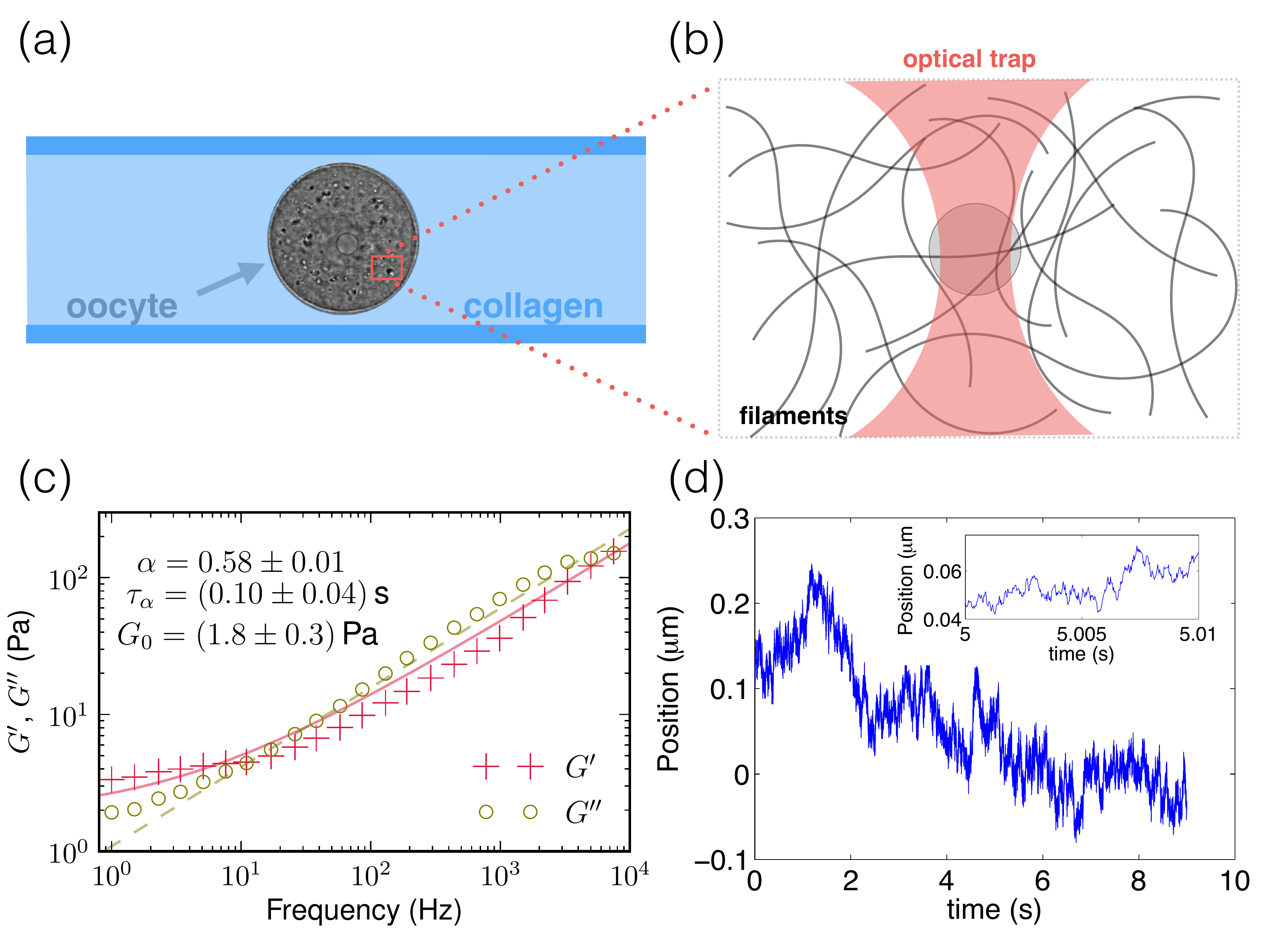}
\caption{\label{fig:sketch}(Color online) Experimental setup for measuring oocyte microrheology. 
	(a)~ We embed oocytes in a collagen matrix between two glass coverslips to prevent rolling during measurement.
	(b)~We use optical tweezers to trap intracellular vesicles and perform active microrheology to measure local mechanical properties~\cite{Mizuno}.
	(c)~Real ($G'$) and imaginary ($G''$) part of the complex modulus, measured from active microrheology. Data at $10$~Hz is used in~\cite{Almonacid}. Best fit curves are in solid and dashed lines for $G'$ and $G''$, respectively.
	(d)~We use laser tracking interferometry to track spontaneous vesicle motion with high spatiotemporal resolution ($10$~nm, $1$~kHz)~\cite{Gittes:1998tg}.
}
\end{figure}

\textit{Experimental setup.}---Mouse oocytes are collected from 13 week old mice and embedded in a collagen gel between two glass coverslips~\cite{Verlhac:1994det}. We measure the local mechanical environment surrounding vesicles in living mouse oocytes using active microrheology~\cite{Mizuno, Ahmed:15b}. We use an optical tweezer to trap vesicles and apply a sinusoidal oscillating force [\fig~\ref{fig:sketch}]. The resulting displacement of the vesicle due to the applied force reflects the mechanical response of the system. We deduce the complex modulus of the intracellular environment surrounding the vesicle from the generalized Stokes-Einstein relation $G^*=1/(6\pi a\tilde{\chi})$, where $\tilde{\chi}$ is the Fourier response function, and $a$ is the vesicle's average radius.

We find that the intracellular mechanics exhibits a power law rheology at high frequencies, and levels off at lower frequencies, as seen in the real and imaginary parts of $G^*$, respectively denoted by $G'$ and $G''$ [\fig~\ref{fig:sketch}(c)]. We fit the experimental data with the function
\begin{equation}\label{eq:G}
	G^*(\omega) = G_0 (1 + (i\omega\ta)^\alp ) ,
\end{equation}
where $\ta$ is a thermal relaxation time scale~\cite{Mason,Wirtz}. To experimentally quantify nonequilibrium dissipation, we must also measure the spontaneous motion of the vesicles by laser interferometry, and extract the power spectral density of the vesicles' position~\cite{Gittes:1998tg}, as is done for passive microrheology~\cite{Mizuno:2008iq} [\fig~\ref{fig:sketch}(d)]. These spontaneous fluctuations entangle information about the thermal and nonequilibrium forces applied on vesicles in the oocyte cytoskeleton~\cite{Ahmed:15b}.


\textit{Caging model.}---We propose a model for the vesicle dynamics and the effect of the surrounding fluctuating actin mesh that takes the observed power law behavior of $G^*$ into account. The model has itself been previously introduced in~\cite{fodor}, but it is generalized here to encompass strong memory effects~\cite{Ahmed:15c}. The underlying physical picture is that the vesicle is caged in the cytoskeleton [\fig~\ref{fig:sketch}(b)], modeled as a harmonic trap of constant $k$, and we assume that nonequilibrium activity induces rearrangements of the cytoskeletal network resulting in a displacement of the cage. In a medium characterized by a memory kernel $\zeta(t)$, we then describe the one dimensional position $x$ of a vesicle with two coupled generalized Langevin-like equations involving the center of the cage $x_0$:
\begin{equation}\label{eq:model}
	\zeta * \f{\dd x}{\dd t} = - k ( x - x_0 ) + \xi , \quad \zeta * \f{\dd x_0}{\dd t} = k \ta \va ,
\end{equation}
where $*$ denotes the convolution product, $\xi$ is a zero mean Gaussian colored noise with correlations $\langle \xi(t)\xi(t')\rangle=\kb T\zeta(\abs{t-t'})$ as provided by the FDT~\cite{Kubo}, and $T$ is the bath temperature. All the degrees of freedom of the surrounding network are embodied by the cage center $x_0$. We have assumed that the dynamics of the network is not affected by the tracer in the regime of our experiment.

The cage motion is given by the active burst $\va$: a zero mean stochastic process representing the random vesicle motion driven by cellular activity~\cite{Almonacid, Ahmed:15c}. In our further analysis, we consider that this process has a single timescale $\tau$ that governs its decorrelation: 
\begin{equation}
 \avg{\va (t)\va(0)} = \f{\kb\Ta}{k\ta} \f{\ee^{-|t|/\tau} }{\tau} ,
\end{equation}
where, by analogy with standard Langevin equation, we have defined an {\em active} temperature $\Ta$ associated to the amplitude of this process. Notice that $\Ta$ is a scalar quantity which quantifies the amplitude of the active fluctuations. We choose the memory kernel $\zeta$ to recover the observed behavior of the measured $G^*$ by adopting a power law decay:
\begin{equation}\label{eq:zeta}
	\zeta(t) = k \pnt{\f{\ta}{t}}^\alp \f{\Theta(t)}{\Gamma(1-\alp)} ,
\end{equation}
where $\Gamma$ is the Gamma function, $\Theta$ is the Heaviside function, and $\alpha<1$. From the generalized Stokes-Einstein relation, we derive that $G^*(\omega)=\brt{k+i\omega\tz(\omega)}/(6 \pi a)$~\cite{mason95}, where the superscript tilde denotes a Fourier transform. The normalization factor in~\eqref{eq:zeta} is chosen so that the corresponding $G^*$ has exactly the same expression as the phenomenological function~\eqref{eq:G} that we use to fit the experimental data, where $k=6 \pi a G_0$. In that respect, the specific form of the memory kernel in~\eqref{eq:zeta} reflects the explicit choice of the best fit curve in~\eqref{eq:G}, yet our approach bears a higher level of generality since it can be extended in a straightforward manner to other kinds of rheology. However, it is not possible to capture the complex frequency dependence of $G^*$ shown in Fig.~\ref{fig:sketch} by discarding memory effects in the dynamics.


\textit{Effective temperature.}---A standard quantification of the departure of the dynamics from equilibrium relies on a frequency dependent ``effective temperature''. Following~\cite{Cug,Nir,Joanny}, it has been introduced by analogy with the FDT as 
\begin{equation}
	\Tf(\omega) = \f{\omega\tC(\omega)}{ 2\kb\tc''(\omega)},
\end{equation}
where $\tC$ and $\tc''$ are the Fourier position autocorrelation function and the imaginary part of the response function, respectively. As a first step, we compute it in terms of the microscopic parameters as
\begin{equation}\label{eq:Tf}
	\Tf(\omega) = T + \f{1}{(\omega\ta)^{3\alp-1}\sin\pnt{\f{\pi\alp}{2}
	}}\f{\Ta}{1+(\omega\tau)^2} .
\end{equation}
The high frequency value collapses to the bath temperature as for an equilibrium behaviour provided that $3\alpha-1>0$. It constitutes a useful benchmark to delineate a thermal regime where active fluctuations are negligible~\cite{Nir}. It also diverges at low frequency as a result of nonequilibrium activity, with a coefficient depending on both the material properties $\{\alp,\ta\}$ and the active temperature $\Ta$. This interplay between mechanics and activity reflects the fact that, in our model, the nonequilibrium processes operating in the system drive motion of the cytoskeletal cage, which in turn affects the vesicle dynamics.


\textit{Dissipation spectrum.}---A quantification of direct physical relevance is the work done by the vesicle on the thermostat~\cite{Sekimoto}, which is the dissipated mechanical energy. The mean rate of energy dissipation $J_\text{diss}$ is the power of the forces exerted by the vesicle on the heat bath, namely the forces opposed to the thermal forces acting on the vesicle by virtue of the action-reaction principle. The thermal forces comprise both the drag force $-\zeta*\dot x$ and the Gaussian noise $\xi$. Therefore, the dissipation rate reads $J_\text{diss}=\avg{\dot x(\zeta*\dot x-\xi)}$, where $\dot x=\dd x/\dd t$ is the vesicle's velocity~\cite{Sasa,Sekimoto}. It is proportional to the rate at which the vesicle exchanges energy with the surrounding environment~\cite{Bustamante}.  In equilibrium $J_\text{diss}$ would vanish, thus expressing the fact that the vesicle releases and absorbs on average the same amount of energy from the thermostat. The dissipation rate is equal to the mean rate of entropy production times the bath temperature $T$. Thereby, it directly characterizes the irreversible properties of the dynamics stemming from the active fluctuations.

The Harada-Sasa equality connects the spectral density $I$ of mechanical energy dissipation to $\tC$ and $\tilde{\chi}''$ in a viscous fluid~\cite{HarSas,harada2006}. It has been generalized to the case of a complex rheology~\cite{PhysRevE.74.026112}, and we express it in terms of the effective temperature as
\begin{equation}\label{eq:I_teff}
	I = \f{ 2 \kb ( T_\text{eff} -T ) }{ 1 + (G'/G'')^2 } .
\end{equation}
This relation allows one to precisely identify the dissipation rate with the nonequilibrium properties of the vesicles' dynamics, since $I$ vanishes at equilibrium. It also enables one to quantify dissipation in the system without any a priori knowledge on the internal source of nonequilibrium fluctuations. The relation~\eqref{eq:I_teff} between effective temperature and dissipation holds independently of our modeling of intracellular activity; it can be used for a large variety of nonequilibrium dynamics. Within our model, the dissipation spectrum is
\begin{equation}\label{eq:I}
	I(\omega)=\f{\pnt{\omega\ta}^{1-\alp}\sin\pnt{\f{\pi\alp}{2}}}{1+2\pnt{\omega\ta}^{\alp}\cos\pnt{\f{\pi\alp}{2}}+\pnt{\omega\ta}^{2\alp}}\f{2\kb\Ta}{1+\pnt{\omega \tau}^2} \,.
\end{equation} 
There is no nonequilibrium dissipation when $\Ta=0$ as expected, while in general it depends on both mechanics and activity as for $\Tf$. By integrating the dissipation spectrum over the whole frequency range, we can deduce the total dissipation rate $J_\text{diss} = \int \dd \omega I (\omega)/(2\pi)$. By contrast to $\Tf$, the dissipation spectrum not only quantifies the deviation from equilibrium properties, it is also related to the energy injected by the nonequilibrium processes.

Using the model~\eqref{eq:model}, the nonequilibrium drive is embodied by the $k x_0$ force applied on the vesicle. The dissipation rate precisely equals the mean power of this force: $J_\text{diss} = \avg{ \dot x k x_0}$, reflecting the fact that the mechanical energy dissipated by the vesicle is also the energy provided by the nonequilibrium processes driving the vesicle's motion. In addition, the dissipation spectrum $I$ equals the Fourier transform of the time symmetric correlation between the vesicle velocity $\dot x$ and the driving force $k x_0$.


\textit{Energy conversion.}---The picture that emerges from our model is that the vesicle motion results from the displacement of the confining cytoskeletal cage, which is due to the active reorganization of the local environment. We denote by $J_\text{env}$ the power of the random force driving the cage's motion. This is the rate of energy injected by the nonequilibrium processes into the environment leading to the cytoskeleton rearrangement in our model. In our model \eqref{eq:model}, it is given by the mean power injected by the force $k\ta\va$ to the cage: $J_\text{env} =  \avg{ \dot x_0 k\ta\va}$, where $\dot x_0=\dd x_0/\dd t$. This can be computed in terms of the microscopic parameters $J_\text{env}=\kb\Ta/\tau(\ta/\tau)^{1-\alpha}$. Note that $J_\text{env}$ can also be regarded as the work per unit time done by the cage on the thermostat, namely on the surrounding medium where the cage is immersed. This interpretation stems from the fact that our model is the limit version of one that features a reaction force of the vesicle upon the cage (for which mechanical interpretations are ambiguity-free), along with thermal fluctuations acting directly on the cage~\cite{fodor2}.

In our phenomenological picture, the energy $J_\text{env}$ injected by the intracellular active processes serves to relentlessly remodel the cytoskeleton network (represented by $x_0$). This energy is then transduced into the vesicles confined in such network (represented by $x$), which is embodied by $J_\text{diss}$, thus driving their active motion. To quantify the efficiency of this energy transduction we introduce the dimensionless ratio $\rho=J_\text{diss}/J_\text{env}$ of the energy  effectively dissipated through active motion of the vesicles over that injected by the nonequilibrium processes into the cage. We find the energetic efficiency $\rho$ to be independent of $\Ta$, and is thus controlled by the time scales $\tau$ and $\tau_\alpha$. We understand such energy transduction as the conversion of the active stirring of the cytoskeleton network into the active dynamics of the intracellular components.


\begin{figure}
\includegraphics[width=\columnwidth]{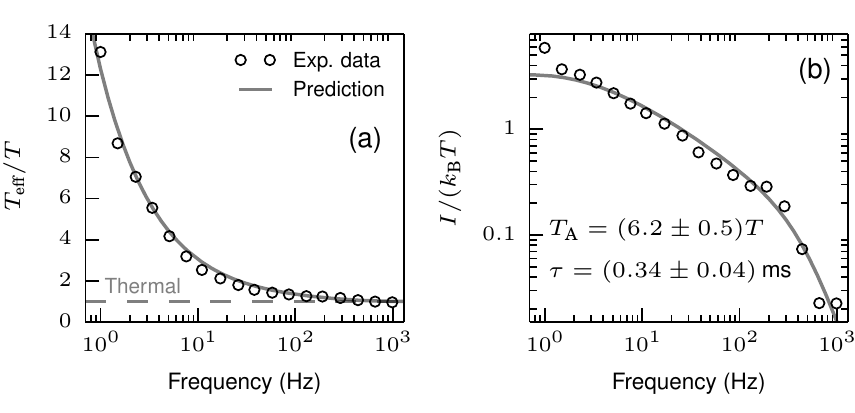}
\centering
\vskip.3cm
\includegraphics[width=.5\columnwidth]{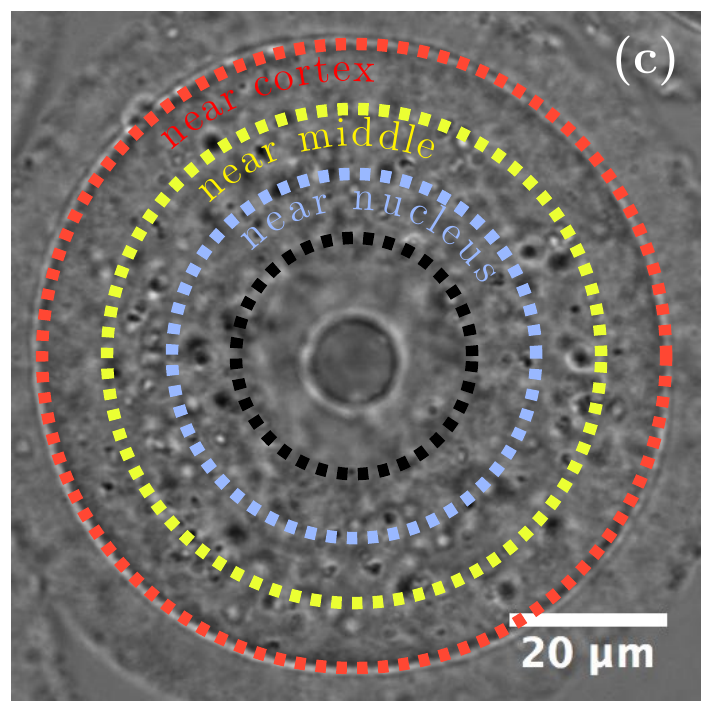}
\vskip.5cm
\includegraphics[width=\columnwidth]{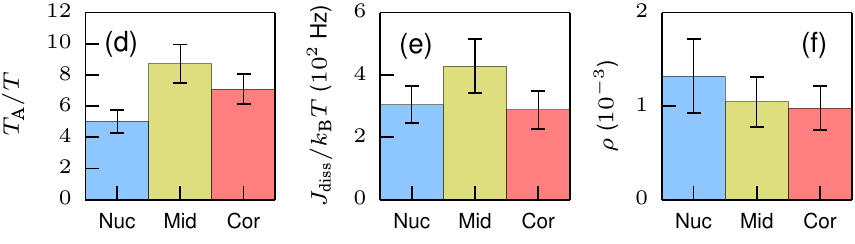}
\caption{\label{fig:diss}(Color online)
	(a)~Effective temperature $\Tf$ and (b)~dissipation spectrum $I$ as functions of frequency ($\circ$), best fits in solid lines using Eqs.~\eqref{eq:Tf} and~\eqref{eq:I}. The horizontal dashed line in (a) is the prediction for a thermal equilibrium system in the absence of activity, for which the dissipation spectrum equals zero.  The experimental data is averaged over the whole oocyte.
	(c)~We distinguish three concentric shells around the central nucleus (black) with a radial extension of about $10$~$\mu$m each: near nucleus (blue), middle (yellow), and cortex (red).
	(d)~Active temperature, (e)~dissipation rate, and (f)~power conversion rate estimated from the best fits of $\Tf$ and $I$ at three locations within the cytoplasm: near the nucleus (Nuc), near the cortex (Cor), and in the region in between (Mid).
}
\end{figure}

\textit{Quantification of the activity.}---We exploit our theoretical predictions to quantify the experimental measurements of nonequilibrium dissipation inside living oocytes. We extract the data for the effective temperature and the dissipation spectrum from a combination of active and passive microrheology. We observe that the experimental effective temperature diverges at low frequency, as a clear evidence that nonequilibrium processes drive the intracellular dynamics in this regime [Fig.~\ref{fig:diss}(a)]. It reaches the equilibrium plateau at high frequency as expected. Deviation from thermal equilibrium was already reported in other living systems~\cite{GalletSM, Betz, Ahmed:15c}. We use our analytic prediction in Eq.~\eqref{eq:I} to fit the dissipation spectrum data.  As we have already quantified the viscoelastic properties, the remaining two free parameters are the ones characterizing the properties of the nonequilibrium processes, namely the active temperature $\Ta$, and the mean persistence time $\tau$. Our best fit is in very good agreement with the measured dissipation spectrum [Fig.~\ref{fig:diss}(b)]. This is the first main result of this paper: our minimal model reproduces very well the available experimental data. It supports the underlying phenomenological picture that the main contribution of the active force driving the vesicle dynamics is mediated by the surrounding network. Moreover, the excellent agreement points to a single dominant active process in the system, characterized by a single time-scale and force magnitude, as captured by the parameters of our model.

The extracted value of the active temperature $\Ta = ( 6.2 \pm 0.5) T $ is larger than the bath temperature $T$. By contrast to $\Tf$, the active temperature is frequency independent, and it quantifies the amplitude of the active fluctuations. Hence, our estimation reveals that the fluctuations due to the nonequilibrium rearrangement of the cytoskeleton have a larger amplitude than the equilibrium thermal fluctuations dominating the short time dynamics. The time scale $\tau = (0.34 \pm 0.04) $~ms that we obtain is of the order of the power stroke time of a single myosin-V motor~\cite{Cappello25092007,Ahmed:15c}. This supports that the nonequilibrium processes driving the vesicle dynamics are related to the microscopic kinetics of the molecular motors. It is consistent with the fact that nonequilibrium processes are dominant at a higher frequency in our system than in others which were mainly driven by myosin-II~\cite{Mizuno,GalletSM}, for which the power stroke time is about $0.1$~s~\cite{CM:CM10014}. It is not an obvious result a priori that one can extract the kinetics of individual motors from global measurements of response and fluctuations. Our analytic prediction for $\Tf$ in Eq.~\eqref{eq:Tf}, for which we use the parameter values $\{\alp,\ta,\Ta,\tau\}$ extracted from the previous fits of $G^*$ [\fig~\ref{fig:sketch}(c)] and $I(\omega)$ [\fig~\ref{fig:diss}(b)], is in consistent agreement with the experimental data [Fig.~\ref{fig:diss}(a)].

To go beyond the calibration of the model presented above, we now use our predictions to investigate energy transfers within the oocytes. From the best fit parameters, we directly estimate the dissipation rate $J_\text{diss}=(360\pm110)$~$\kb T/\text{s}$, as well as the power conversion rate $\rho=(1.7\pm0.8)10^{-3}$. The definition of $J_\text{diss}$ is independent of any modeling of the underlying activity, whereas $\rho$ depends on our specific model. We find that the conversion of energy from the cytoskeletal network to the vesicle is very low. This is the second main result of this paper. It suggests that a major proportion of the nonequilibrium injected power is dedicated to the network rearrangement, and not necessarily to vesicle dynamics \textit{per se}. In other words, the injected energy tends to go mostly into elastic stresses, and only a small fraction ends up in kinetic energy~\cite{Gov:14, Gov:15}. Note that the power of the active force driving the vesicle being small compared with the one moving the cage is not in contradiction with our assumption that the vesicle driving is mediated by the cage. Indeed, such a driving already leads to a significant deviation from equilibrium, as shown in Fig.~\ref{fig:diss}(a).

It has been reported that a single myosin-V motor does about $3$~$\kb T$ of work during one power stroke~\cite{Fujita}, from which we deduce that it dissipates approximately $10^4$~$\kb T/\text{s}$ into the intracellular environment during the power stroke. This result is to be compared with our estimation of $J_\text{env}=(2.0\pm0.5)10^5$~$\kb T/\text{s}$. We infer that the power injected by the nonequilibrium processes into the environment represents approximately the activity provided by $20$ myosin-V motors. Assuming that the nonequilibrium processes in oocytes are indeed mainly regulated by myosin-V activity, we infer that $20$ is the typical number of motors involved in the nonequilibrium reorganization of the cytoskeletal cage in the vicinity of a vesicle.


\textit{Variability across the oocyte.}---One of the main advantages of our energetic approach lies in the ability to compare the same physical quantities  across a large variety of living systems, or in different locations of the same system. We consider three concentric shells within the oocyte cytoplasm located near the nucleus, near the cortex, and in between these two regions. Each shell has a radial extension of about $10$~$\mu$m [Fig.~\ref{fig:diss}(c)]. We fit the real and imaginary parts of $G^*$ for the three regions, and we use our analysis to quantify the corresponding $\Ta$, $J_\text{diss}$, and $\rho$. Our results hint that nonequilibrium activity is increased near the middle of the cell, and slightly decreased near the nucleus, as quantified by $\Ta$ and $J_\text{diss}$ [Figs.~\ref{fig:diss}(d-e)]. This suggests that living oocytes locally regulate the nonequilibrium activity throughout their cytoplasm by injecting different amounts of energy. Note that the relative variation of $J_\text{diss}$ and $\Ta$ are similar, showing the close relation between these quantities as highlighted in Eq.~\eqref{eq:I}.

In comparison, the variation of $\rho$ does not exhibit a clear trend across the oocyte [Fig.~\ref{fig:diss}(f)]. Since the $\rho$ definition is a balance of a purely active parameter $\tau$ and the material properties $\{\alpha, \tau_\alpha\}$, this result suggests that the nonequilibrium fluctuations are adapted in each region to the local mechanical properties. It is known that the molecular motors do not only produce active forces in the cell, they also affect its mechanical properties~\cite{Ahmed:15c}. Therefore, we speculate that there might be a feedback between the overall fluctuations induced by the motor activity and the mechanics of the surrounding cytoplasmic network within which they move to find an optimum rate of power conversion, the optimal value being roughly the same for the three locations.


\textit{Conclusion.}---We quantified the amount of mechanical energy dissipated by the intracellular dynamics. Our analysis utilizes a minimal model describing the effect of the nonequilibrium stochastic forces in living systems with complex rheology. We find the predictions of our model to be in excellent agreement with the experimental results for vesicles in living mouse oocytes, thus allowing us to quantify the main properties of the nonequilibrium dynamics: the amplitude and typical time scale of active fluctuations, the amount of dissipated energy, and the rate of energy transmitted from the cytoskeletal network to the intracellular components. The extracted parameters provide a quantitative support to the experimental picture that the nonequilibrium processes are mainly driven by myosin-V activity~\cite{Howard,Schuh,Almonacid, Ahmed:15c}. The use of general principles in stochastic energetics, together with a minimal microscopic model, makes the results of our study highly relevant to a large variety of nonequilibrium processes in biology and active matter~\cite{Chaudhuri:13, Chaudhuri:14, Nardini:16, Nardini:16b}.

\acknowledgments{We warmly thank Gavin Crooks for a critical reading of the manuscript and FvW acknowledges the support of the UC Berkeley Pitzer Center for Theoretical Chemistry. WWA thanks the PGG Fondation and Marie Curie Actions.}


\bibliographystyle{eplbib}
\bibliography{mapoo-references}

\end{document}